# Dynamics of airflow in a short inhalation

A. J. Bates[1], D. J. Doorly[1], R. Cetto[1,3], H. Calmet[4], A. M. Gambaruto[4], N. S. Tolley[3], G. Houzeaux[4] and R. C. Schroter[2]

[1]Department of Aeronautics, and [2]Department of Bioengineering, Imperial College London, London SW7 2AZ, UK
[3]Department of Otolaryngology, St Mary's Hospital, Imperial College Healthcare Trust, London W2 1NY, UK
[4]Computer Applications in Science and Engineering, Barcelona Supercomputing Center (BSC-CNS), Barcelona 08034, Spain

AJB, 0000-0002-8855-3448; DJD, 0000-0002-5372-4702; RC, 0000-0002-7681-7442;
GH, 0000-0002-2592-1426

During a rapid inhalation, such as a sniff, the flow in the airways accelerates and decays quickly. The consequences for flow development and convective transport of an inhaled gas were investigated in a subject geometry extending from the nose to the bronchi. The progress of flow transition and the advance of an inhaled non-absorbed gas were determined using highly resolved simulations of a sniff 0.5 s long, $1 \, l \, s^{-1}$ peak flow, 364 ml inhaled volume. In the nose, the distribution of airflow evolved through three phases: (i) an initial transient of about 50 ms, roughly the filling time for a nasal volume, (ii) quasi-equilibrium over the majority of the inhalation, and (iii) a terminating phase. Flow transition commenced in the supraglottic region within 20 ms, resulting in large-amplitude fluctuations persisting throughout the inhalation; in the nose, fluctuations that arose nearer peak flow were of much reduced intensity and diminished in the flow decay phase. Measures of gas concentration showed non-uniform build-up and wash-out of the inhaled gas in the nose. At the carina, the form of the temporal concentration profile reflected both shear dispersion and airway filling defects owing to recirculation regions.

## 1. Introduction

Inhaled air traverses a series of bends and changes in passage area as it passes via the nose to the early generations of the bronchi. Throughout this region, shown in figure 1 for the particular subject anatomy considered, the speed of the air and the width of the passages vary greatly. In most parts, the Reynolds number of the flow is at least in the several hundreds, so that a wide gallery of airflow patterns occur, influencing local physiological function.

Although healthy human airways are broadly similar, regional patterns of flow differ according to subject geometry, as shown in many studies [1–3], though the complex anatomy makes exploring the relation using analytical descriptions [4] difficult. The temporal dynamics also affect the flow behaviour, as shown, for example, by the experimental studies of Chung & Kim [5]. The unsteady aspect of airflow is less explored, because the time scale of a breath is sufficiently long during quiet, restful breathing that flow inertia does not excessively dominate viscous forces. More specifically, the Womersley parameter, $\alpha$ (defined as $\alpha = R(\omega/\nu)^{1/2}$, where $R$ is a length scale, $\omega$ is the angular frequency and $\nu$ is the kinematic viscosity), is near enough to 1 that airflow may be assumed quasi-steady.

In high-frequency ventilation of the trachea [6,7], large values of $\alpha$ occur, but this is an artificial condition, very different from normal breathing. However, even for normal breathing cycles, differences between quasi-steady and unsteady simulations have been observed. Hörschler *et al.* [8], for example, found significant hysteresis in the pressure–flow relationship in accordance with the low-frequency flow variation during the breathing cycle.

Rapid acceleration of the flow during inhalation increases the frequency content of the flow waveform and thus the value of $\alpha$; Rennie *et al.* [9] found the time taken for flow to build up to a level equivalent to that in restful breathing was 130 ms on average in normal inspiration. This reduced to 90 ms in sniffing, and was as short as 20 ms in a rapid inhalation.

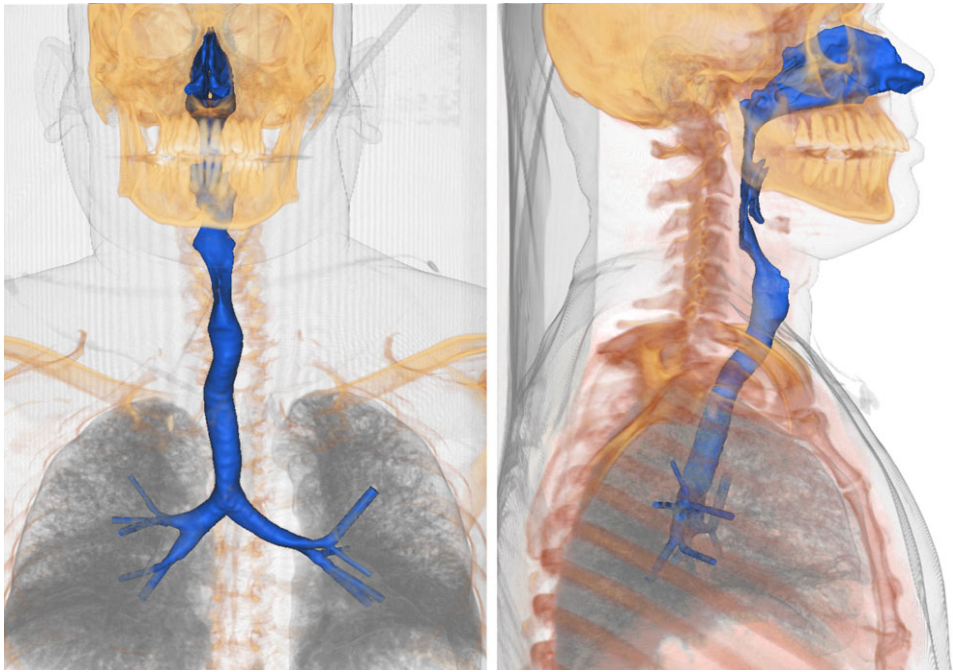

**Figure 1.** Volume rendering of a neck and chest CT scan with the segmented airway surface highlighted. A stereo lithography file of the surface is provided in the electronic supplementary material. (Online version in colour.)

Air flow in a rapid inhalation is of physiological interest to understand sniffing for olfaction or during manoeuvres commonly recommended when taking aerosol medication. It also reveals the time scales over which flow responds.

The inhalation flow profile used was defined by fitting a sixth-order polynomial (see appendix A) to experimentally measured data from a previous *in vivo* investigation [9]. The graph of total airflow against time is shown in figure 2, which also shows the flow through each nostril. The subject paused before the start of inhalation and at the end, so flow commences from an initial state different from that expected of a cyclic breathing pattern. This is an important distinction from cyclic breathing, as are the depth of inhalation and rapidity of acceleration [9].

The spatial and temporal resolution of the computations were set at a high level to fully resolve the flow dynamics and provide data to serve as a benchmark for further computational investigations of airway flow dynamics. The airflow simulations were obtained by solving the unsteady Navier–Stokes equations with mesh parameters such that the average ratio of mesh length scale to Kolmogorov length scale was 5 (better than 2–3 in the areas of local refinement), whereas the time step of $1.0 \times 10^{-5}$ s was smaller than the Kolmogorov time scale in over 99% of interior cells; such spatial and temporal resolutions should fully capture the flow dynamics. More details of the computational methodology are given in appendix A.

Previous studies have been constrained by the available technology, limiting airflow models to steady flow [10,11] or to allow only slow temporal variations [12]. Subsequent improvements in computing and imaging facilitated study of flow initiation in a unilateral nasal model [13], with continuous particle image velocimetry measurements obtained at an effective rate of 15 000 fps validating direct numerical simulations (DNSs). The development of a prominent flow recirculation in the anterior portion of the nasal cavity was mapped, as was the instability of the shear layer at the margins of the jet issuing from the nasal valve, though only in the initiation phase and for a single side.

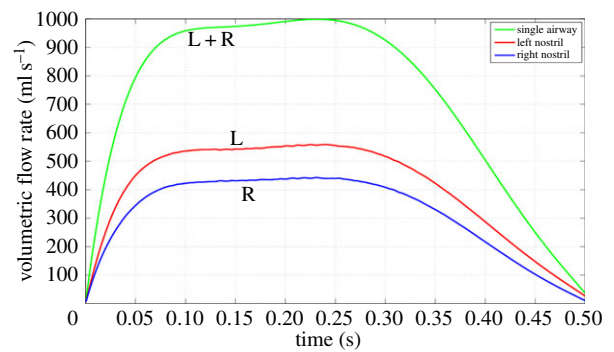

**Figure 2.** Volumetric flow rate showing differences between left and right sides. (Online version in colour.)

Ishikawa *et al.* [14] simulated a sniff where the flow peaks at a slightly lower flow rate than that in the present investigation (around 800 ml s$^{-1}$) in one nasal cavity, but at a relatively coarse numerical resolution, whereas the main focus of their analysis concerned flow in the olfactory groove.

The distribution of flow into various regions in the nasal airways is of considerable interest, both in normal breathing and in sniffing. The question has been investigated by both Zhu *et al.* [1] and Segal *et al.* [15], who performed steady-state simulations at lower breathing rates, representing restful breathing rather than sniffing. Higher nasal velocities have been modelled previously [16], achieving speeds 50% higher than this study in the nasal jet. The study [16] reported that, for steady breathing rate simulations, 3.5 million mesh elements are required for converged pressure loss predictions, but this is for modelling assumptions that do not resolve the detailed unsteady internal dynamics, namely the spontaneous fluctuations associated with transitional or turbulent flow. Until relatively recently, these features were modelled by

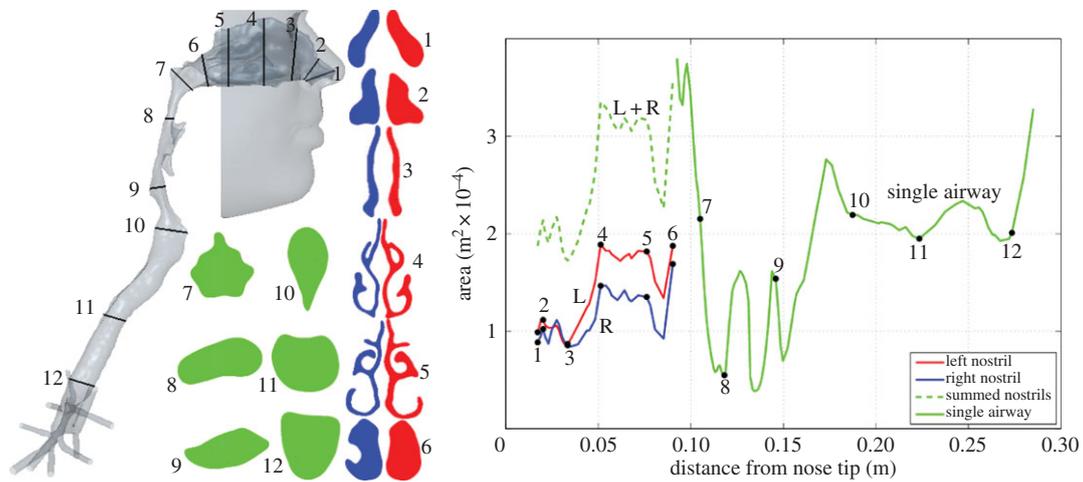

**Figure 3.** Geometry of the airway with flow-normal cross-sectional areas. In the nose, the area distributions are shown separately for the left and right cavities. (Online version in colour.)

Reynolds averaged Navier–Stokes approaches. For example, Ghahramani *et al.* [17] investigated turbulent intensities and their effect on particle deposition in an extensive airway model ranging from the nostrils to above the carina at flow rates comparable to the peak flow rates used in this study. Further studies [18,19] of particle deposition reported four million cells as necessary for mesh independent results. However, small-scale fluctuations would not be sufficiently resolved at this refinement level, nor are the effects of accelerating and decelerating flows covered in steady simulations.

Zhao *et al.* [20] compared models of sniffing using laminar simulations and $k-\omega$ and Spalart–Allmaras turbulence models and found little difference in particulate absorption between the models.

The dynamics of spontaneous flow instability were investigated by Lin *et al.* [2], who reported simulations of a realistic oral geometry. Large-scale flow features were resolved, whereas a comparison between an under-resolved DNS and a large eddy simulation turbulence model was found to have a relative L1 norm (sum of absolute differences in a volume as a percentage of the total) of turbulent kinetic energy (TKE) of 13.6%. This highlights the need for simulations at higher resolutions to serve as reference studies.

Zhang & Kleinstreuer [21] and Pollard *et al.* [22] investigated the flow in a highly idealized oral airway model. The former study detailed the TKE at flow rates of 10 and $30\,\mathrm{l\,min^{-1}}$. High values of TKE were found in the region below the epiglottis. Various computational models of turbulence were compared with experimental results, with high-resolution simulations concluded as desirable.

## 2. Material and methods

### 2.1. Anatomy

The airway geometry was derived from a neck and chest computed tomography (CT) image set of a 48-year-old man in the supine position. Note that anatomical left and right are used throughout: figures labelled subject right are shown on the left and vice versa. The images were enhanced with contrast and were collected retrospectively from a hospital database. The scan was requested for clinical reasons and reported to be normal. Consent from the patient for subsequent use of the data was obtained. Further details of the subject, data acquisition and translation to a virtual geometry are given in appendix A.

### 2.2. Geometry definition for simulation

The face was placed inside a hemisphere of radius 0.5 m and flow imposed uniformly on the curved outer surface of the hemisphere, allowing a natural inflow to develop. (The consequences of prescribed inflow at the nostril have been described previously [23,24].) The flat plane of the hemisphere was located approximately 0.05 m behind the chin and treated as a rigid wall; limiting the exterior airspace has negligible effect as little inhaled air originates behind the head [23]. To provide a well-defined outlet, the final bronchial branches were extruded into constant cross-section pipes for several diameters.

The graph in figure 3 shows the cross-sectional area of the airway, with the nostrils marked as plane 1 and plane 12 lying just above the carina, i.e. where the trachea divides. The minimum cross-sectional area in the nose occurs at plane 3 in the diagram (area $1.82 \times 10^{-4}\,\mathrm{m^2}$, corresponding hydraulic diameter of $4.94 \times 10^{-3}\,\mathrm{m}$). This plane is located slightly posterior to the nasal valve, often assumed to be the minimum area [25,26]. In the postseptal airway, a series of minima occur in the laryngopharyngeal zone in the region of planes 8 and 9. The local area increases between these minima correspond to openings such as that into the back of the mouth and are not representative of the area used by the bulk of the flow; the effective flow area is closer to the minimum shown at plane 8. The glottis (plane 10) does not correspond to the minimum area airway. This is a consequence both of the vocal cords being abducted and of the supine position of the subject, which results in the tongue and soft tissues reducing the airway area about planes 8 and 9.

The left and right nasal cavity airspaces are separately delineated in figure 3. The left nasal airspace has a volume of $14\,600\,\mathrm{mm^3}$, slightly larger than on the right, $12\,300\,\mathrm{mm^3}$. The associated surface areas are, however, similar, yielding a surface area-to-volume ratio for the left nasal airway of $0.717$ and $0.827\,\mathrm{mm^{-1}}$ for the right, mean $0.768\,\mathrm{mm^{-1}}$. The nasal airway volumes were measured from the nostril to a plane which extends the floor of the nasal cavity through the nasopharynx; the nasal volumes were also split in the post-septal region by another plane which is tangential to the posterior aspect of the septum. This differs from the method employed by Segal *et al.* [15], who used a coronal plane at the back of the septum as the posterior marker for the nasal volume. The latter volume does not necessarily include the posterior aspect of the inferior turbinate, the size of which can have a significant impact on airway volume.

Figure 4 shows the position of planes selected to illustrate the dominant features of the flow.

### 2.3. Boundary conditions

The time-varying inflow velocity was applied to the external hemisphere, whereas the ends of the extended bronchi were held at constant pressure throughout. A no-slip condition was applied to the flat surface of the hemisphere and to all anatomical surfaces. No significant differences were found when (i) the radius of the hemisphere representing the external far-field boundary was increased from 0.5 to 1 m from the centre of the face or (ii) the simulation was re-run using a prescribed pressure as an inlet condition rather than specifying the velocity. The walls are considered to be rigid, as Fodil *et al.* [27] demonstrated that there is no significant change in area until higher nasal pressure drops than are considered here, i.e. greater than 100 Pa.

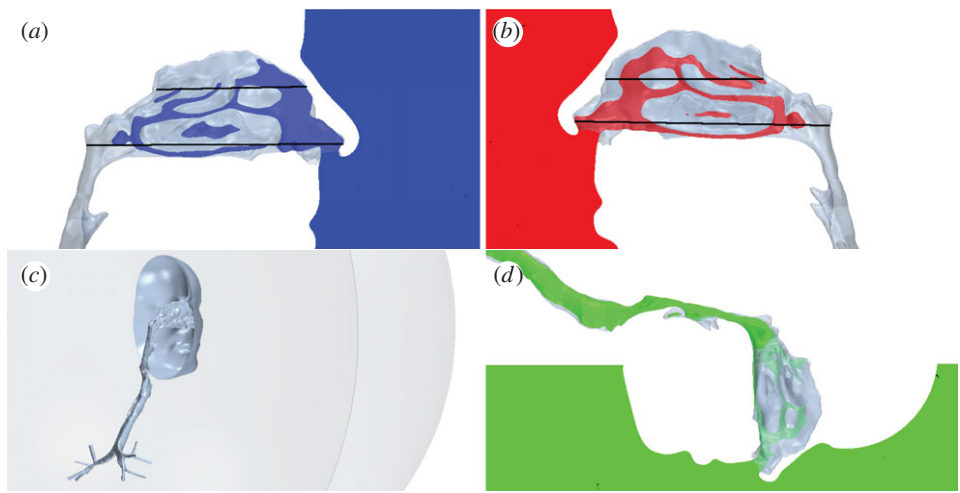

**Figure 4.** Positions of planes within the airways: (*a*) sagittal through right nasal cavity; (*b*) sagittal through left nasal cavity; (*c*) the position and extent of the exterior hemispherical boundary; (*d*) sagittal through the laryngopharyngeal and glottal regions. (*a*,*b*) show the positions of axial planes within the nose. (Online version in colour.)

## 3. Results

### 3.1. Flow development

Inhaled air divides naturally between the nostrils, with that entering the left nostril (figure 2) approximately 1.25 times greater than on the right (figure 2) throughout the inhalation. Integrating the flow curves yields a total inhaled volume of 364 ml (204 ml through the left nostril and 160 ml through the right). Between 0.1 and 0.25 s, flow approximates a plateau, with corresponding Reynolds numbers in the nose (based on the hydraulic diameter of the minimum cross-sectional area (plane 3)) of approximately 1900 on the right and approximately 2400 on the left; by contrast, in the supraglottic region (plane 9), $Re$ is approximately 14 000. Figure 5 shows the nasal flow field on sagittal planes during the ramp-up phase (at four instants) and near flow cessation.

After 10 ms elapsed time (figure 5*a*) flow is at low speed (less than 3 ms$^{-1}$) and attached everywhere, resembling potential flow.

After 20 ms (figure 5*b*), flow recirculation is developing in the upper anterior portion of the right cavity (note left side of figure 5*b*), posterior to the nasal valve; it appears similar to that observed in the study of Doorly *et al.* [13] for a different geometry. No such recirculation is, however, apparent in the left airway. Both sides of the nose display a rapid increase in passage height after the nasal valve, but the abruptness of the increase is greater on the right, inducing flow separation.

After 50 ms, the flow pattern appears fully established in each cavity. Little further change, except in velocity magnitude, appears in the image pair at 100 ms (figure 5*d*). In the anterior region of the right cavity, the separation occupies a significant volume, acting to 'block' the ingress of newly inhaled air. Flow in the left side is somewhat faster, particularly in the region of the jet formed through the nasal valve. Flow from the nasal valve appears less uniform on the right with smaller regions of fast-moving flow interspersed with slower regions. Other features evident include smaller recirculation zones in the nostril vestibules and low flow in the portions of the inferior meatus cut by the image planes. The final image pair (figure 5*e*) corresponds to 0.45 s, when the total flow rate has diminished to approximately 250 ml s$^{-1}$, about the same level as that in the first image pair at 0.01 s (figure 5*a*). The anterior recirculation zone has begun to enlarge and continues to grow as flow decays further; on the left, flow is simply reduced in magnitude.

The evolution of regional airway resistance, defined as the total pressure difference divided by flow rate, is compared for the transnasal airspaces (planes 1–6) and the descending airway (planes 7–12) in figure 6. In steady flow, the resistance depends only on flow rate. The inhalation profile considered involves significant flow acceleration up to 0.1 s, which requires an additional transient pressure gradient. At 0.05 s, the left transnasal pressure loss is approximately 39 Pa, but rises to approximately 50 Pa at 0.1 s. However, the resistance is virtually unchanged, as the component needed for an acceleration that is diminishing happens to be balanced by that needed to drive an increasing flow. This balance is coincidental. The same does not occur in the descending airway, which is the dominant resistance for the subject, owing to the reduced area in the oro- and hypopharynx. The raised resistance in this region is consistent with the supine position of the subject, as Anch *et al.* [28] found this position to significantly increase supraglottic resistance, whereas Cole & Haight [29] found nasal resistance unchanged. It would appear that the pharyngeal soft tissues are responsible in this case, but to what extent this is unusual is not known.

In a separate investigation, a constant inlet flow rate equal to that at 0.13 s was instantaneously applied. After only approximately 0.02 s, airway resistance and flow partitioning metrics throughout approached an equilibrium level of fluctuation. An instantaneous flow start shortens the initial transient and, while unphysiological, may be of interest as a procedure to predict airway resistance at a fixed flow rate using direct numerical simulation at lower computational cost. These quasi-steady simulations lack the correct convective history and, although they may be suitable for the coarse measures above, may not be appropriate for measures such as scalar uptake or particle transport.

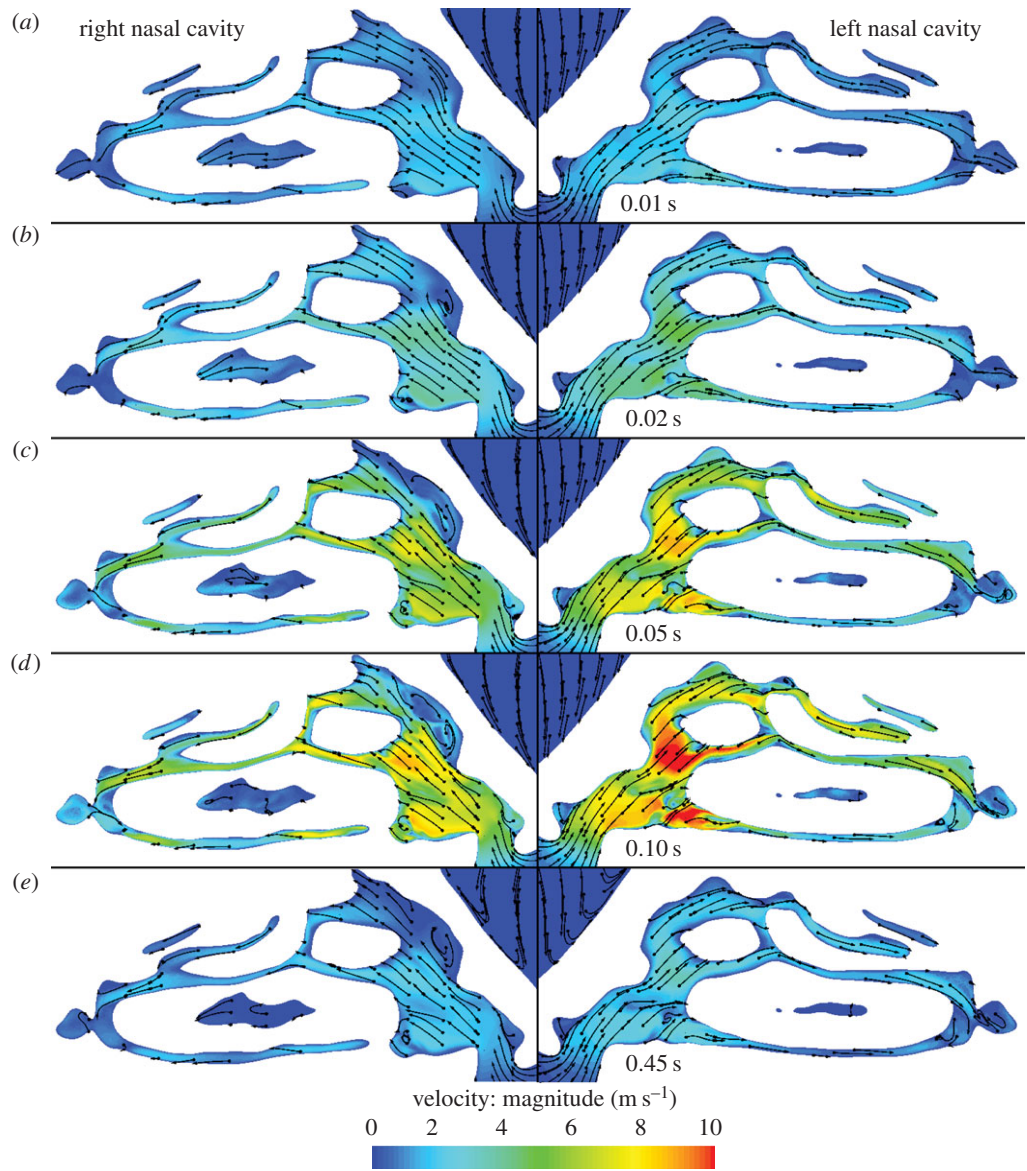

**Figure 5.** (a–e) Velocity contours and planar streamlines on two planes in the right and left nasal cavities at succeeding instants.

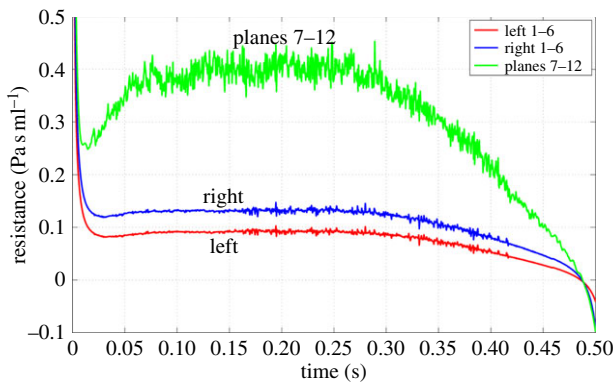

**Figure 6.** Comparison of transnasal and descending airway resistances. (Online version in colour.)

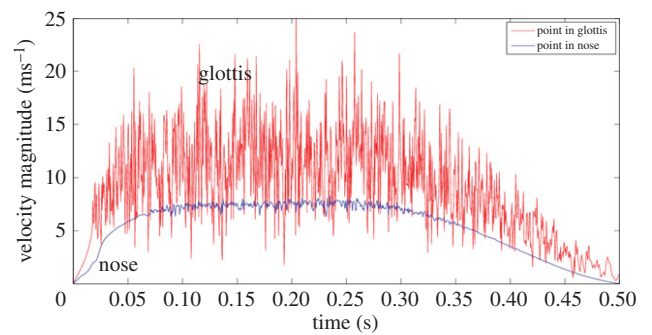

**Figure 7.** Flow velocity signals at two points: lower curve—point in the right nasal cavity towards the top of the intersection of the right sagittal plane (figure 4) and plane 3 (figure 3); upper curve—point in the glottis at the intersection of the sagittal plane (figure 4) and plane 10 (figure 3). (Online version in colour.)

Returning to the sniff, spontaneous fluctuations associated with turbulence develop earlier in the glottic region than in the nose, in line with the near order of magnitude greater Reynolds number. This is evident in figure 7, where samples of velocity magnitude are compared for a point on plane 10 within the glottal region with those from a point on the sagittal plane in the right nostril, halfway between planes 2 and 3. In the nose, flow speed increases calmly until 0.07 s (corresponding flow rate approx. 900 ml s$^{-1}$) then oscillations of relatively small amplitude appear; these diminish as flow decays and are absent after 0.35 s (flow rate approx. 750 ml s$^{-1}$).

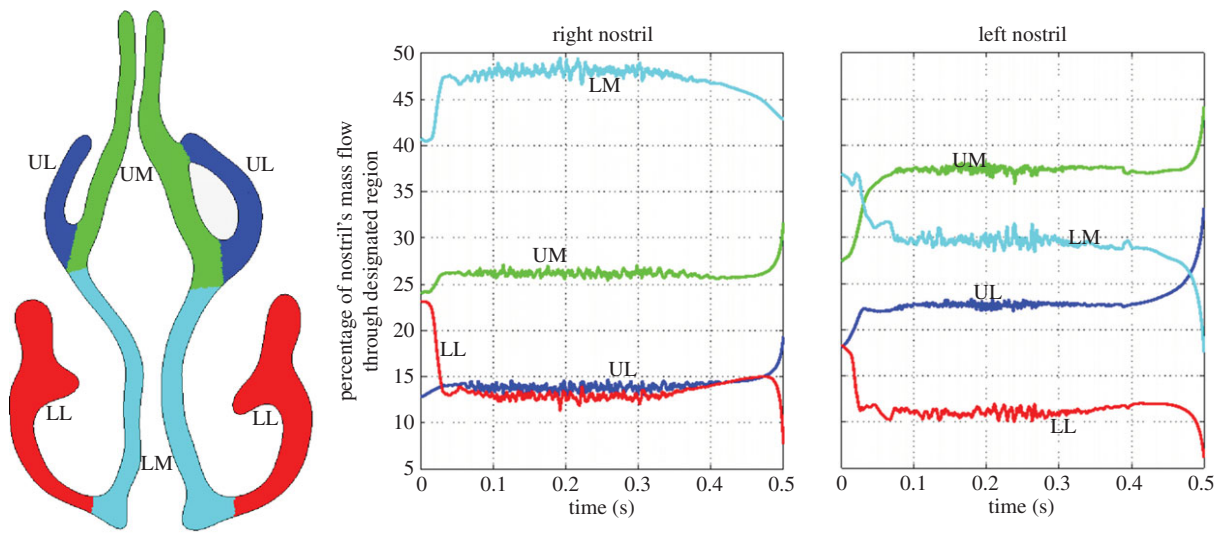

**Figure 8.** Mid-nasal cavity (figure 3, plane 4) regional division and corresponding flow distribution. UM, upper medial; LM, lower medial; UL, upper lateral; and LL, lower lateral. (Online version in colour.)

Flow in the laryngeal region becomes unsteady after 0.02 s, by which time the local mean airspeed reaches around 6 ms$^{-1}$. There the oscillations persist, merely reducing somewhat towards flow cessation.

## 3.2. Patterns of flow: nasal flow division

The architecture of the nose partitions flows into separate streams to best accomplish physiological functions such as heat and water exchange [30] and sampling of inhaled air by the olfactory receptors. Airflow partitioning during inhalation was quantified by computing the flux through different sectors of nasal plane 3 (figure 3).

Plane 4 lies in the turbinate region and is subdivided into sectors referred to as the upper and lower medial and lateral quadrants in figure 8. The division between the upper and lower regions was affected by positioning a plane orientated in the axial direction between the middle and inferior turbinates, so that the plane did not touch either structure. Asymmetries in the turbinates on either side of the nose slightly displace this division on the left side relative to that on the right.

In figure 8, the graphs indicate how the partitioning of flow in each cavity varied during the inhalation. Three distinct stages in the dynamics of the flow split are apparent, which we describe as initiation, equilibrium and decay.

In the right cavity, after approximately 0.03 s, the flow split equilibrates. On the left, it appears to take almost twice as long. In both cases, it is somewhat surprising that the equilibrium flow balance is established quickly, despite the flow having attained only approximately 50% of its peak and still accelerating. However, most of the gross changes in flow pattern occur within this period as shown in figure 5. The initiation process sees the potential flow-like pattern change once the nasal valve jet develops and recirculation zones appear where the flow separates. The time scales for these events depend on the length scales of local features; in this part, the time is broadly comparable to the 50 ms required for the inhalation to fill a volume equivalent to that of the nose.

As flow initially accelerates, the jet entering the nasal cavity strengthens, diverting more flow to the upper regions and around the middle turbinate rather than along the floor of the nasal cavity. This accounts for the rapid rise of flow to the upper regions on both sides of the nose during the establishment phase.

Minor fluctuations in the proportions of flow directed through the various sectors reflect local unsteady disturbances at peak flow rates. The manner in which the flow streams are divided differs between left and right sides, despite the division of the nasal airspace yielding quadrants of proportionately similar size on the left and right. On both sides, the bulk of the flow passes along the medial side of the respective cavity, i.e. along the septal wall; this accounts for 73% of the flow at equilibrium in the right nostril and 66% in the left. However, the divide between the flow transiting upper versus lower regions is different: on the right 60% of flow favours the lower regions, whereas on the left 60% favours the upper. This seems to be linked to the flow pattern illustrated in figure 5, where the recirculation zone in the right nostril prevents flow to the upper regions, causing a bias of flow to the lower regions. With no separation in the left nostril, flow appears more evenly spread between the various divisions.

Deceleration of the flow commences after 250 ms, but the flow divisions do not change until later, perhaps 400 ms. Even then there is little change until the final few tens of milliseconds when certain features developed in the initial phase of the flow alter. The inertia of the nasal jet continues to direct air to the upper reaches of the nose, increasing those regions' share of flow.

## 3.3. Patterns of flow: unsteadiness and turbulence

Figure 9 shows the mean velocity, $V_{mag}/V_{ref}$, and the fluctuating velocity, $V'_{mag}/V_{ref}$, across two sagittal planes, one in each nostril, whereas corresponding quantities for the supraglottic region are shown in figure 10. $V'_{mag}$ is defined as the square root of the variance of the velocity magnitude and is proportional to the square root of TKE. These quantities are calculated over a 30 ms window of the sniff, starting at 0.1 s, when flow approaches a plateau. The reference velocity taken for normalization in the nose ($V_{ref,nostrils} = 3.52$ ms$^{-1}$) is obtained by taking the spatial and temporal mean of velocity on plane 4 (figure 3) during the 30 ms window.

Mean velocities appear higher and more uniform in the left nasal cavity than in the right. The right cavity shows regions of fast-moving flow negotiating the obstacles of the septal wall and turbinates, with a separation region in the anterior portion of the right nostril. Plots of $V'_{mag}$ also reveal flow on the right to be more unsteady than on the left, other than at the head of the middle turbinate in plane 4 where the septal spur is located. It is noteworthy that, in the nose, regions of significant fluctuation appear localized and quite different between the two sides.

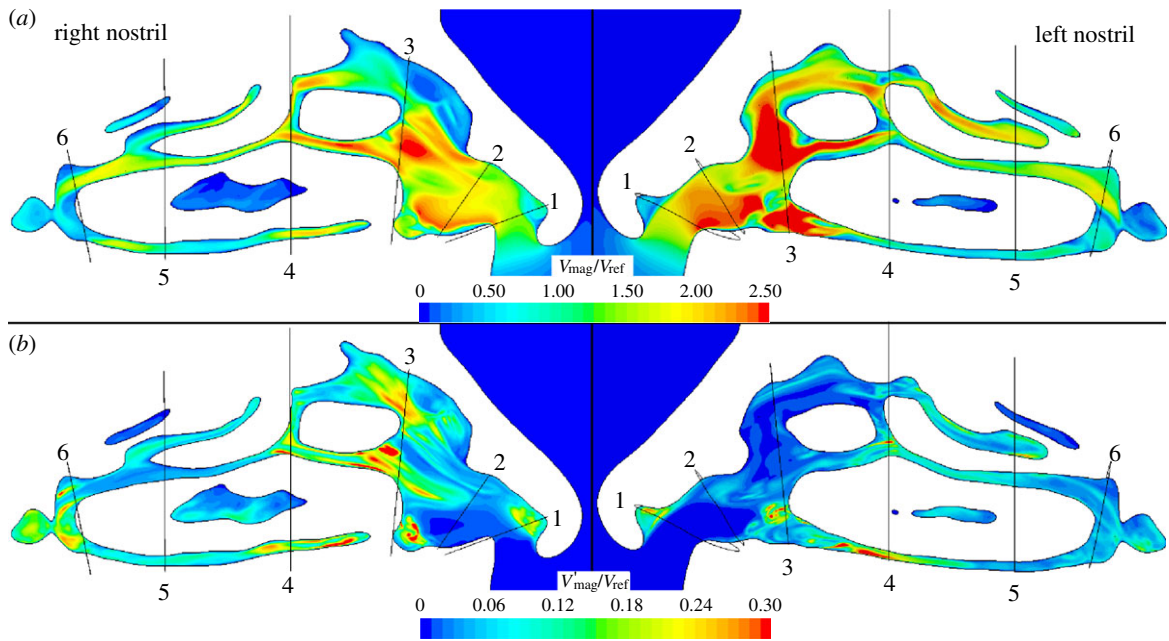

**Figure 9.** (a) Mean $V_{mag}$ and (b) $V'_{mag}$ in the nose at peak flow, scaled by local reference velocities.

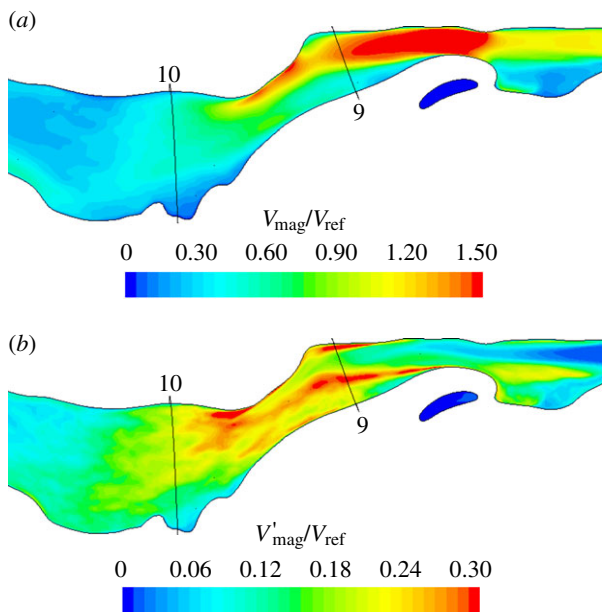

**Figure 10.** (a) $V_{mag}$ and (b) $V'_{mag}$ in the oropharynx, hypopharynx and larynx at peak flow, scaled by local reference velocities.

The exterior flow is sink-like and laminar in nature. The flow in the left cavity remains largely laminar in the anterior region, although there is evidence of transitional flow on the right side, especially around the breakdown of the nasal-valve jet (where the sagittal plane crosses plane 3).

Flow in the cross-sectional planes 3 and 4 in the nose and at planes 9 and 10 in the descending airway is compared in figure 11. Comparing left and right cavities at plane 3, in the upper region of the right cavity $V_{mag}$ is low, whereas $V'_{mag}$ is high, owing to the unsteady recirculation in this zone. Plane 4, positioned at the head of the middle turbinate and slightly posterior to the septal spur, reveals some unsteadiness in flow on the left. It can also be observed that whereas the right cavity flow is concentrated in the lower medial section, along the septal wall, flow on the left is more uniformly distributed with highest velocities around the head of the middle turbinate. Downstream of this cross section, the flow in the nose appears calmer, as the passage traversed is more uniform in shape until the nasopharynx.

Flow in the supraglottic region (figures 10 and 11) is dominated by the jet issuing from the constrictions through the epiglottis and the back of the tongue. Here, the normalizing value, $V_{ref} = 17.7 \text{ ms}^{-1}$, corresponds to the average velocity at plane 8. At plane 9, flow effectively bypasses much of the available airway area. The peak fluctuation intensity can be seen in the plots of figures 10 and 11 to occur at the margins of the jet. Direct comparison of the flow in this region with idealized problems, such as the mildly eccentric stenosis considered in Varghese et al. [31], is difficult, given that the airway is a more complex shape and the inflow is highly disturbed. Visualizations of lambda 2 isosurfaces, however, show streamwise vortical structures in the jet at its narrowest point just upstream of plane 9, with some evidence of hairpin-like vortical structures in the vigorous breakdown at plane 9. The asymmetry of the jet is still visible by plane 10, where the fast-moving flow is directed along the subject's left-hand side.

Elevated values of $V'_{mag}$ at plane 10 point to the continued high fluctuation amplitude as the jet impinges on the side wall. By plane 11 (not shown), the fluctuations are much reduced as the flow fills the trachea more uniformly. Summary values of the fluctuation intensity integrated over the various cross-sectional planes are shown in table 1, comparing the magnitude of velocity fluctuations and the turbulence intensity, defined as the ratio of a representative fluctuation velocity to the mean flow velocity.

The rich detail of the flow that lies behind the unsteadiness is revealed by plotting the instantaneous magnitude of vorticity (defined as the curl of the velocity field) in the nose and in the supraglottic region (figure 12). The arrow in the right nasal cavity (figure 12a) points to break-up of the shear layer at the margin of the anterior recirculation. The pair of arrows in the supraglottic region (figure 12b) highlights successive shear layer instabilities, demonstrating the complex inflow condition for the downstream jet.

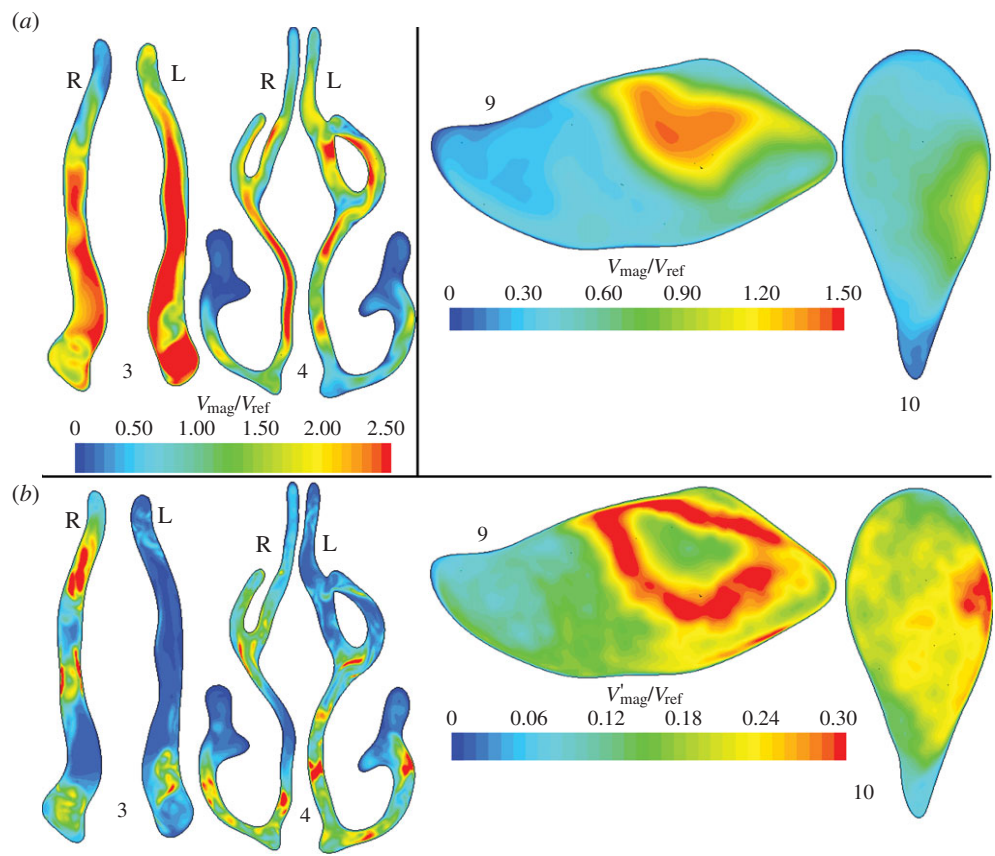

**Figure 11.** Through-flow (a) $V_{mag}$ and (b) $V'_{mag}$ velocities in anterior and mid nasal cavity (figure 3, planes 3 and 4) and supraglottic and glottic regions (figure 3, planes 9 and 10).

**Table 1.** Mean values of velocity and turbulence quantities on planes defined in figure 3 in the period 0.1–0.13 s.

| plane | $V_{mag}$ (ms$^{-1}$) | $V'_{mag}$ (ms$^{-1}$) | $T_u$ (%) |
|---|---|---|---|
| 1 (L–R) | 5.6 – 5.5 | 0.11 – 0.13 | 4 – 3 |
| 2 (L–R) | 6.4 – 5.9 | 0.06 – 0.10 | 1 – 1 |
| 3 (L–R) | 7.2 – 5.8 | 0.13 – 0.39 | 1 – 7 |
| 4 (L–R) | 3.5 – 3.4 | 0.30 – 0.31 | 9 – 9 |
| 5 (L–R) | 3.5 – 3.5 | 0.22 – 0.17 | 5 – 5 |
| 6 (L–R) | 3.6 – 3.2 | 0.20 – 0.50 | 5 – 5 |
| 7 | 4.9 | 0.36 | 6 |
| 8 | 17.7 | 0.47 | 2 |
| 9 | 10.8 | 3.37 | 22 |
| 10 | 8.8 | 3.39 | 24 |
| 11 | 5.3 | 0.88 | 11 |
| 12 | 5.2 | 0.64 | 8 |

Figure 12 points to the breakdown of separated shear layers as the primary cause of fluctuations in airway flow. The fluctuation energy is far larger in the airway through the supraglottic and glottic zones than in the nose, but is of a different character. This is shown by plotting the temporal energy spectra of the velocity fluctuations in the nose and supraglottic region in figure 13, at the points indicated by crosses in figure 12. Given the limited duration of the temporal window, the spectra are processed in the manner referred to in Varghese *et al.* [31], but using just two Hann windows. Figure 13 shows the (unnormalized) fluctuation energy in the supraglottic region to be more than two orders of magnitude greater than in the nose; moreover, the velocity spectrum at the point in the nose lacks the broadband energy content of that in the downstream regions. Normalizing the spectrum indicates the emergence of a limited inertial subrange, consistent with $Re_\lambda$ of around 20 [32], computed at the point.

## 3.4. Transport by the flow of inhaled scalar

The fate of inhaled species depends on several parameters, but principally their size and mass. Larger particles do not follow path lines of the flow owing to factors such as size, shape, inertia and gravitational effects. These effects diminish with particle size, though additional displacements owing to Brownian motion become significant at very small scales. Pure convective transport by the flow is more approximately realized for nanoparticles, which are small enough to neglect inertial effects, but sufficiently larger than molecular species to limit diffusion. To investigate the relation between the dynamics of airflow and convective transport during rapid inhalation, a scalar species with low diffusivity (Schmidt number = 900) was introduced and tracked. The advantage of this choice of species parameters (negligible absorption, minimal diffusion) is that it identifies characteristic minimum times for species to be transported to the varying parts by convection alone, free of the complex confounding effects when absorption and diffusion are simultaneously varied. Further work will consider the effects of different parameters such as diffusion, surface absorption characteristics and the nature of the inhaled species. At all non-outlet boundaries, the scalar concentration gradient was held at zero, to prevent absorption by the surface, whereas the outlets afforded no barrier to escape. Tracking this idealized scalar reveals how the air flow field directs inhaled species to enter and to fill the various regions of the airway in the absence of species-dependent absorption and diffusion characteristics.

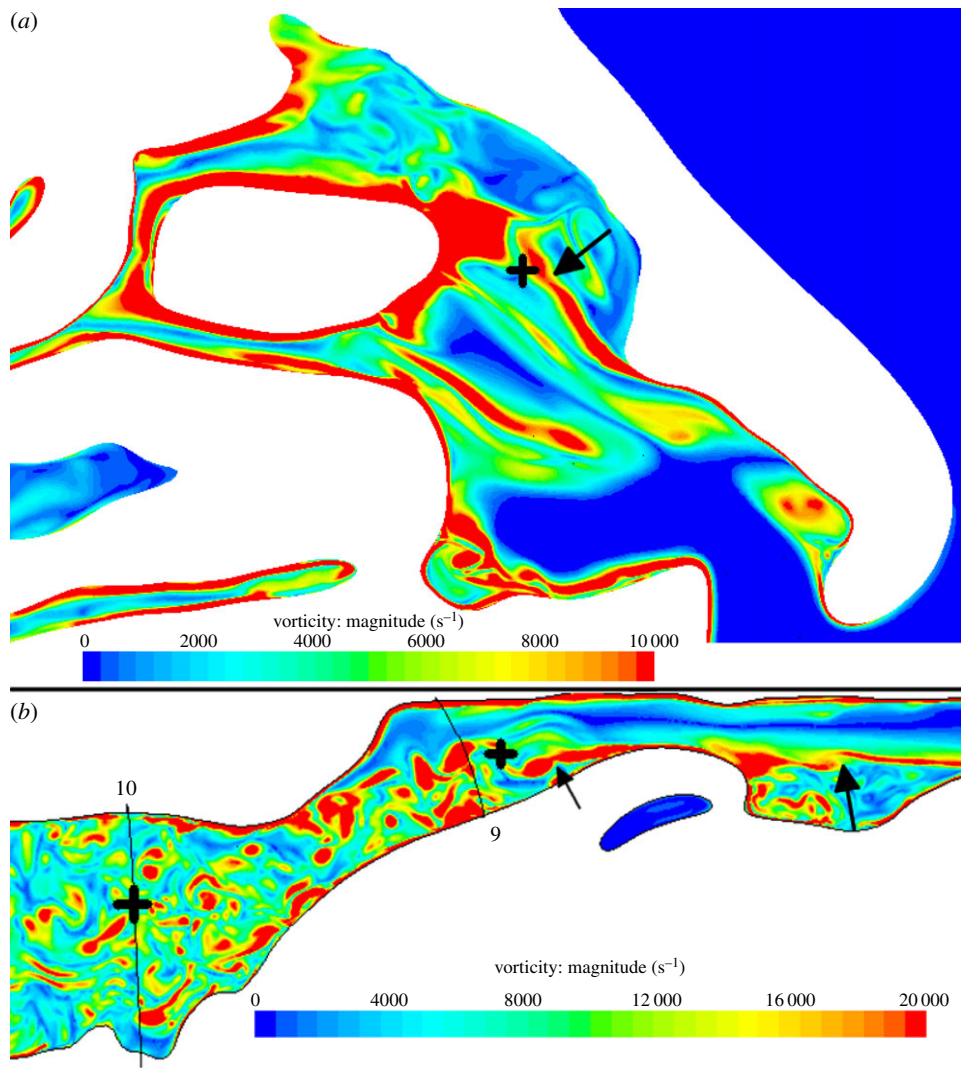

**Figure 12.** Contours of vorticity magnitude on sagittal planes in the right nasal cavity (*a*) and oropharyx, hypopharynx and larynx (*b*). The arrows illustrate shear layer instabilities. The black crosses illustrate the points at which the spectra in figure 13 are calculated.

Before the inhalation, the scalar was confined to an approximately hemispherical volume of 364 ml (the total volume inhaled) positioned against the face, with the widest point (diameter) level with the nostrils (figure 14).

Figure 15 shows the initial progress of the scalar into each nasal cavity, where shading is used to mark the 50% concentration threshold at various instants up to 0.1 s.

Figure 15 shows how the local recirculation regions that develop hinder the scalar advance. This recirculation zone is similar to that found experimentally by Schreck *et al.* [33], Kelly *et al.* [34] and Doorly *et al.* [35] and inhibits scalar ingress into the upper regions of the nose (where olfactory receptors are located most densely) throughout the early part of the sniff. As noted by Kelly *et al.* in comparing their work with earlier studies by Schreck *et al.*, the size and occurrence of recirculation zones are geometry dependent; here, there is no comparable recirculation in the left cavity.

Between 0.01 and 0.02 s, the scalar front has progressed beyond the vestibule and at 0.03 s has reached the head of the middle turbinate. In addition, at 0.03 s, the scalar has progressed further into the left cavity, where velocities are higher than in the right.

On both sides, the scalar progression is greater above than below the inferior turbinate. By 0.04 s, almost the entire left cavity is flooded with scalar, whereas, on the right, the white, unfilled region marks the recirculating air in the upper anterior cavity.

After 0.05 s, the scalar field has reached the back of the turbinates and is entering the nasopharynx region. Although not evident in the sagittal plane shown, the scalar takes a similar amount of time to reach the olfactory cleft located in the upper region of the nasal cavities.

After 0.1 s, both cavities appear well filled with the notable exception of the recirculation regions and dead-end passages, which are bypassed by the bulk flow. Overall, the pattern of scalar advance agrees well with in-plane velocities in the nose, owing to the relatively narrow calibre of the passages.

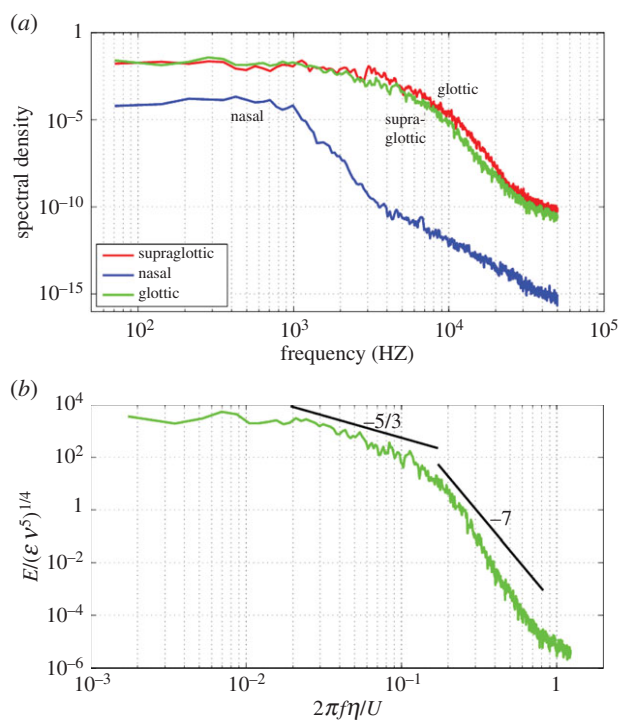

**Figure 13.** (a) Summed power spectra of fluctuating components of velocity at points in the nose (lower curve), supraglottic region (upper curve) and glottic region (middle curve). (b) Spectra from the point in the glottis, scaled by the viscous dissipation rate, $\varepsilon$, and the kinematic viscosity, $\nu$. The frequency, $f$, is scaled by the $2\pi\eta/U$, where $U$ is the mean velocity at that point and $\eta$ is the Kolmogorov length scale. (Online version in colour.)

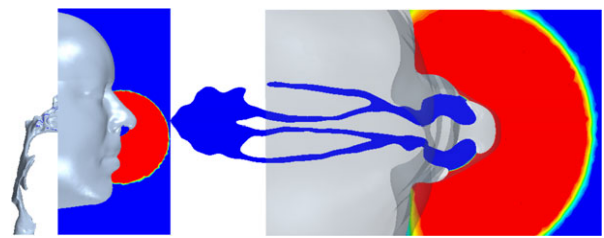

**Figure 14.** Initial scalar distribution.

Figure 16 shows the advance of the scalar in the external and internal domains over the entire course of the inhalation. Scalar concentration is shown on the lower axial plane described in figure 4. At 0.04 s, saturated scalar has filled the nostrils and spread either side of the inferior turbinate in each cavity. Mixing by the complex velocity field dilutes the scalar front, which is entering the nasopharynx, carried by a pair of post-septal jets. By 0.1 s, the nasal cavity is almost saturated.

The effect of the physical separation of the nostrils can be seen in the external scalar distribution at 0.25 s. Whereas, earlier, the scalar appears drawn to a single sink, the external scalar front is now retreating towards the nostrils more rapidly at the sides. Hence, at 0.25 s, scalar-free air begins to be drawn in from the sides and below the nostrils, before all the external scalar has been inhaled.

By 0.5 s, the scalar has almost disappeared externally, with 81% of the available scalar already inhaled. Some scalar remains within each cavity, with less in the right nostril as this has been washed out by clean air owing to high velocity in the lower medial region (figure 8).

The extent of dispersion of the advancing scalar front, resulting from the combination of narrow passageways, flow division and recombination throughout the nose, is revealed in figure 17, where the scalar concentrations on the higher axial plane (figure 4) and on planes 1–7 from figure 3 are shown at 0.04 s.

The imprint of the anterior recirculation in the upper right cavity on the scalar field is again evident in both views, corresponding to a zone of low scalar concentration. Similarly, recirculation regions affect the anterior region of each nasal vestibule in plane 1.

At plane 4, the bulk of the flow bypasses the upper parts of the inferior meatus, which remains devoid of scalar at this time. At plane 5, differences between the flow in each cavity are also apparent in the more homogeneous distribution of scalar on the left side; the relative concentration distributions accord with the partitioning of airflow shown in figure 8.

Planes 6 and 7 show that the greater flow in the left cavity results in a larger, but less intense post-septal jet, whereas the effect of the mixing of jets in the nasopharynx on scalar distribution can be seen in plane 7.

The arrival of scalar and its decay as fresh air is inhaled was computed for each of the planes shown in figure 3 with the results from four planes shown in figure 18.

The shape of these curves approximately follows that of the volumetric flow profile shown in figure 2, although it must be noted that the concentration of inhaled scalar begins to become diluted from about 0.1 s (figure 18). Thus, the scalar flux through plane 1 shows a slight downwards slope to the plateau, contrary to the slight increase in volumetric flow beyond 0.1 s; because the initial dilution is small, the effect is very slight. By plane 4, this effect is offset somewhat by recirculation regions filling through mixing with the main flow.

The time taken for the scalar to first appear at successive downstream planes increases from virtually zero at the nostril plane to 0.03 s at the head of the middle turbinate (plane 4) and 0.07 s at the carina (plane 12). This latter time is actually shorter than the time taken to inhale a quantity of air equivalent to the volume of airway down to the carina, 0.093 s. The difference between these times indicates that newly inhaled air does not effectively fill the entire volume, whereas the non-constant velocity profile causes some scalar dispersion.

Integrating curves of scalar flux shows that 292 ml of scalar was inhaled out of a total inhaled volume of 364 ml. Of the inhaled scalar, 265 ml passed plane 12, the carina entering the bronchial tree, whereas 27 ml remained in the airways. Some scalar (10 ml) was left in the nasal region (between planes 1 and 7), representing a concentration of 40% in the nasal volume.

Figure 19 shows the saturation of the scalar species in the upper region of the nostril planes, defined as that above the upper axial plane shown in figure 4. This region includes the olfactory cleft, so scalar intrusion into this region makes sensory detection more likely. Both nostrils fill their upper regions to above 95%, showing that, despite the recirculation zones and low flow areas, the mixing of the flow is sufficient to flood both nostrils eventually. The left cavity fills more quickly than the right, owing to the higher flow rates on this side, whereas on the right the recirculation zone fills slowly through mixing with the main flow. These effects mean that scalar clears the right cavity more slowly.

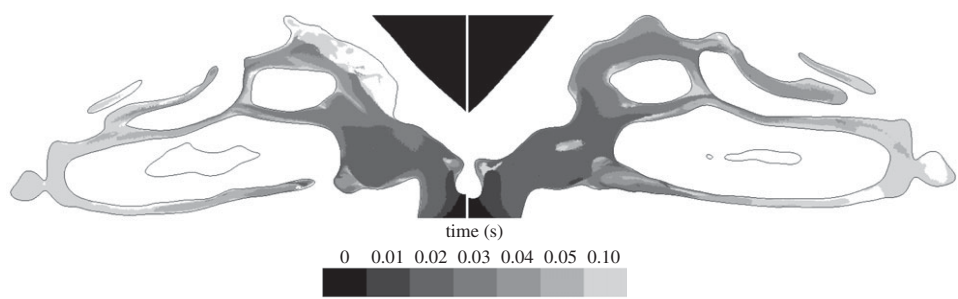

**Figure 15.** Planar contours of 50% scalar concentration at successive instants (coded by shading) in the nasal cavity.

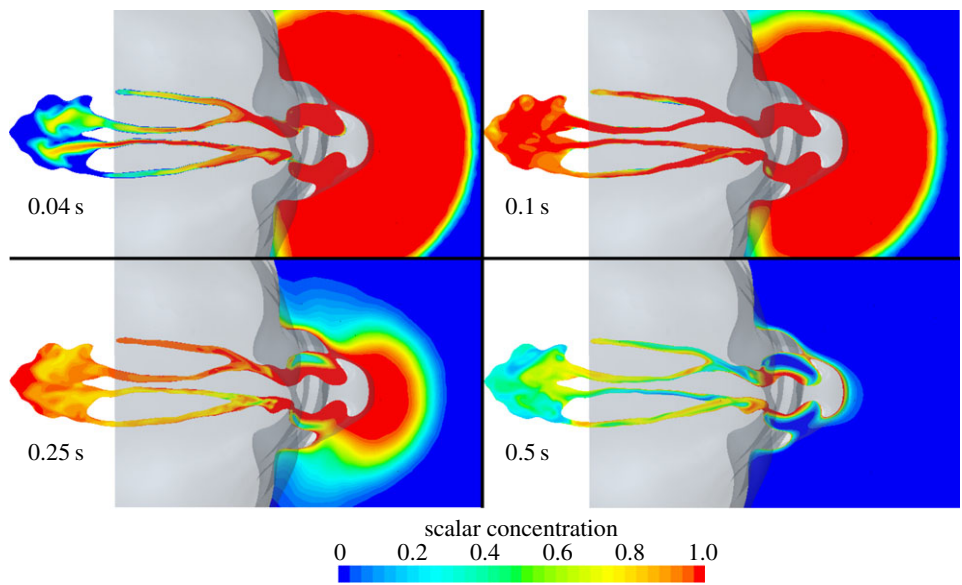

**Figure 16.** Distribution of scalar species on the lower axial plane (figure 4).

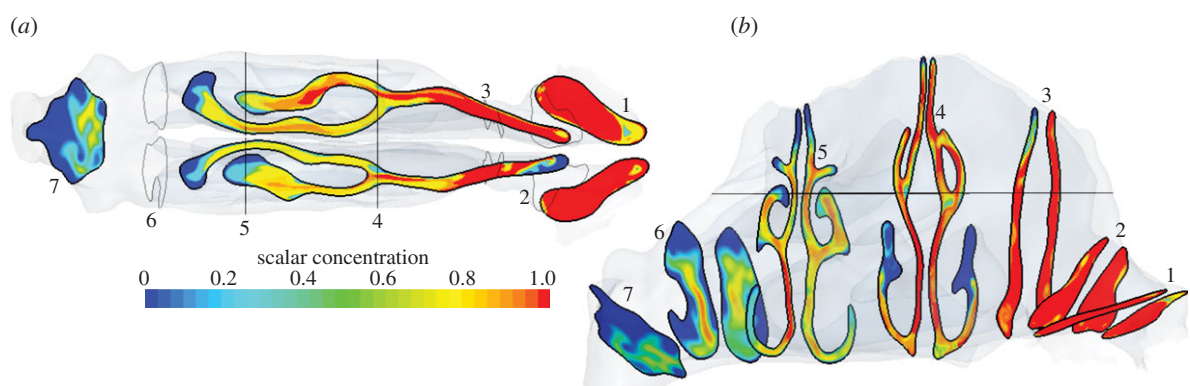

**Figure 17.** Instantaneous scalar distribution in the nose at 0.04 s on (a) planes 1−7 (figure 3) and (b) the upper axial plane (figure 4).

## 4. Discussion and conclusions

Numerical simulation was applied to chart the dynamics of the airflow and the transport of an inhaled gas during a rapid inhalation such as a sniff. The inhalation profile had overall parameters: 0.5 s duration, $1 \, 1 \, s^{-1}$ peak flow and inhaled volume 364 ml, derived from previous *in vivo* measurements by an analytical fit. A realistic airway geometry was used extending from the nose to the large bronchi, whereas computations employed high spatial and temporal resolutions (cell volume of order $0.001 \, mm^3$ in refined regions, time step $1.0 \times 10^{-5} \, s$).

Local flow patterns strongly affect convective transport throughout the airways, as shown by tracking the ingress and partial clearance of a non-absorbing scalar of low diffusivity. The patterns of flow changed both spatially in response to the airway cross-sectional area and shape, and temporally as the flow rate built up and decayed. Significant differences in flow structure and flow distribution were found between the left and right nasal cavities. Where the airway enlarged rapidly in the nose and in the supraglottic region, separated shear layers were produced, becoming unstable when the local velocity became sufficiently high. In the supraglottic region, fluctuations of a relatively high amplitude developed early in the acceleration phase and persisted to the end of the inhalation. In contrast, fluctuations in the nose developed later and died out as the flow diminished below 70% of the peak.

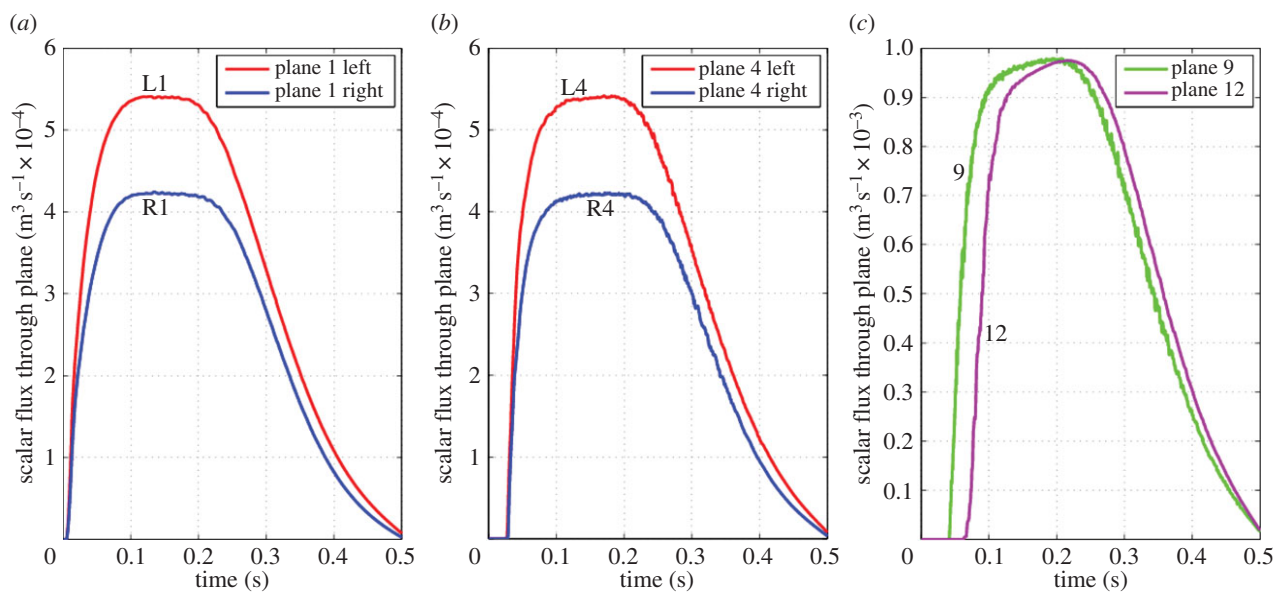

**Figure 18.** ($a$–$c$) Scalar flux $\int \phi \mathbf{v} \cdot \hat{\mathbf{n}} \, da$ through the nostril (figure 3, plane 1, left), mid-nasal cavity (figure 3, plane 4, centre), supraglottic region (figure 3, plane 9, right) and the distal trachea (figure 3, plane 12, right). ($a$,$b$) show the two sides of the nasal cavity separately. (Online version in colour.)

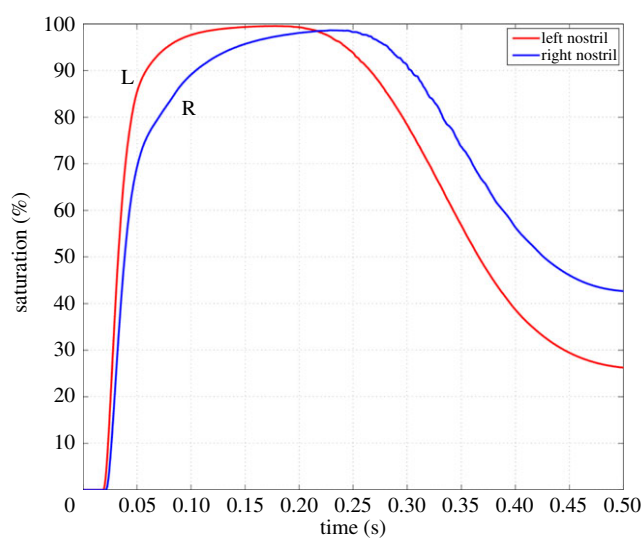

**Figure 19.** Scalar build-up and decay in the nasal region above the upper axial plane (figure 4), calculated as $\int \phi \, dV$ and expressed as a percentage of the total volume of the upper region in each nasal cavity. (Online version in colour.)

The dynamics of the flow development followed significantly different time scales in the various zones; consequently, it is inappropriate to assume that flow behaviour simply scales with the inhalation profile. Key features of the observed evolution of inspiratory flow can be summarized as follows.

(1) In the nose, different patterns of flow appeared in three phases, labelled initiation, quasi-equilibrium and decay. The intracavity partitioning of nasal flow altered part way through the initiation or flow build-up phase. This corresponded to the establishment of localized flow separations and was of similar duration to the time required for the initial flow and the scalar to fill the nasal cavity (50 ms). Thus, flow initiation was essentially completed for the nose significantly before peak inhalation flow.

(2) During the approximately flat plateau phase of the bulk of the inhalation, the partitioning of flow between various parts of each cavity, and between cavities, was virtually constant, save for minor disturbances owing to flow-field fluctuations.

(3) Spontaneous fluctuations in local flow velocity were observed at an early stage (approx. 20 ms) in the region of the glottis, whereas fluctuations in the nasal cavity were delayed nearly to peak flow (approx. 100 ms) and were of a lower amplitude. High turbulence intensities (fluctuating velocities of around $10 \, \mathrm{m \, s^{-1}}$) occurred at the margins of the jet in the supraglottic region, whereas, in the nose, fluctuations were lower (less than $1 \, \mathrm{m \, s^{-1}}$) and occurred predominantly in the right cavity. The pattern of flow transition, both spatial and temporal, differed not just along the airway but between both sides of the nose.

(4) Transport of the scalar to the carina by the flow was achieved rapidly. Near saturation of scalar concentration at the carina was attained approximately 100 ms later than at a mid-point in the nasal cavity.

(5) Local flow patterns are capable of significantly affecting scalar transport, as shown by the emergence of a prominent anterior recirculation in the right nasal cavity, which acted to delay the scalar advance in that area.

(6) Finally, in the decay phase, the partitioning of flow between the parts of the nasal cavities remained unaltered until approximately the final 100 ms (in some parts even later). From that time on during the decay period, a greater concentration of scalar was retained in the upper right nasal cavity than in the left, indicating slower clearance from the right side.

The complexity of the detailed flow properties observed in this study is of significance to modelling and quantification

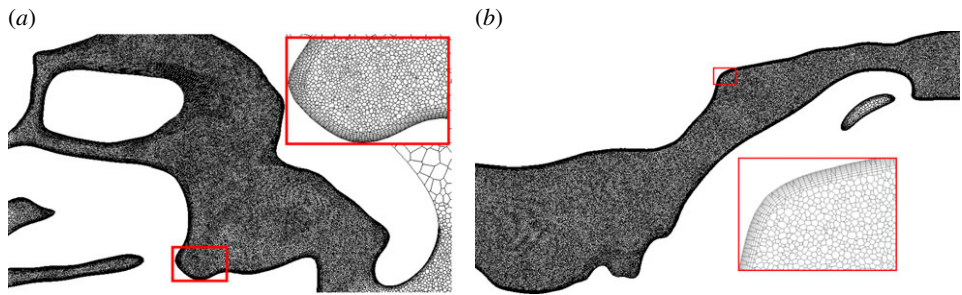

*(a)* *(b)*

**Figure 20.** Details of the airway mesh on planes in the nose (*a*) and oropharynx, hypopharynx and larynx (*b*). (Online version in colour.)

of the transport of both gaseous and particulate reactive and passive species within the nose. In relation to this, in this work, we are charting the spatio-temporal development of the low- and high-frequency components of the flow and examining the effects of varying diffusivity and surface absorption characteristics particularly on scalar transport. Future work should also investigate the effects of repetitive sniffs and scalar transport in combined inspiratory and expiratory flow manoeuvres.

Funding statement. This research was supported by EPSRC Doctoral Training Award EP/P505550/1.

# Appendix A

## A.1. Data acquisition

The data were acquired using a Philips-Brilliance 64 scanner and consisted of 912 slices of 1.0 mm thickness with a pixel size of $0.65 \times 0.65$ mm. A consultant radiologist reported the nasal airways as clear and of normal appearance. The position of the tongue base and other soft tissues in the pharynx were deemed consistent with the patient being scanned in the supine position. The vocal cords were noted to have the appearance of being abducted, whereas the trachea was considered to be of normal dimensions, not demonstrating any abnormalities. The airway in the pharynx may be narrower than if the patient had been standing, but the geometry is within the normal range.

The airways were determined by segmentation whereby an ENT surgeon extracted the region of interest from the CT image stack and created a three-dimensional surface outline using AMIRA v. 4.2 (FEI Visualization Sciences Group). A small angulation of just over 3 mm of the bony nasal septum is seen projecting into the left cavity at the level of the inferior turbinate in the form of a septal spur. The resulting surface was then smoothed using Taubin's [36] algorithm in MESHLAB (Visual Computing Laboratory). The smoothed surface was then merged with the external non-anatomical surfaces to define the boundaries of the flow domain.

## A.2. Mesh creation and analysis

The airspace was meshed using Star-CCM+ 8.04.007-r8 (CD-adapco). The mesh elements were polyhedral in the bulk of the geometry with prismatic layers lining the walls. Fourteen million elements were used. The choice of mesh element is significant as comparisons between polyhedral and tetrahedral meshes have shown that polyhedrons can resolve the flow to the same resolution as a tetrahedral grid with up to five times as many elements [37]. This is especially true for complex geometries containing recirculation regions, as they are less sensitive to alignment with the flow direction, reducing the advantages of structured meshes. The geometry was divided into regions, with cells more densely allocated in the regions in which the flow was more disturbed. In regions such as the anterior nasal airways and the larynx, nine prism layers were used with the first element from the surface having a height of approximately 0.02 mm. For cells crossing planes 2 and 3, as described in figure 3, the mean volume was $2.7 \times 10^{-3}$ mm$^3$. All cells within the airways had a volume of less than $1.0 \times 10^{-2}$ mm$^3$, although they could be much larger in the external geometry, especially far from the face. Figure 20 shows a cross section through the mesh in two of the refined regions, (*a*) the right nostril and (*b*) the glottis region.

The $y_+$-value is less than 1 throughout the internal geometry. In the refined regions of the nose, it has a mean value of 0.38 and across planes 9 and 10 (figure 3) the mean is 0.77. An equivalent to the $y_+$-value can be defined in the bulk of the flow, using the Von Mises yield criterion instead of wall shear stress to calculate a friction velocity and the cube root of cell volume rather than first cell height for the length parameter. This calculation produces mean values of 2.5 in the nose and 5.2 across plane 9.

The temporal discretization of the simulation was set at $1.0 \times 10^{-5}$ s. This yielded mean $\Delta t_+$-values of 0.22 in the nose and 0.92 across plane 9. $\Delta t_+ = \Delta t \, u_\tau^2/\nu$, where $u_\tau$ is the friction velocity, $\Delta t$ is the time step and $\nu$ is the kinematic viscosity.

## A.3. Flow metrics

The turbulent dissipation rate, $\varepsilon$, was directly computed from the gradients of the velocity field (equation (A 1)) [38] and a temporal average taken over the selected window, in order to evaluate local values of the Kolmogorov and Taylor scales,

$$\varepsilon = \nu \left\{ 2 \left( \overline{\left( \frac{\partial u'_1}{\partial x_1} \right)^2} + \overline{\left( \frac{\partial u'_2}{\partial x_2} \right)^2} + \overline{\left( \frac{\partial u'_3}{\partial x_3} \right)^2} \right) \right.$$
$$+ \overline{\left( \frac{\partial u'_1}{\partial x_2} \right)^2} + \overline{\left( \frac{\partial u'_2}{\partial x_1} \right)^2} + \overline{\left( \frac{\partial u'_1}{\partial x_3} \right)^2}$$
$$+ \overline{\left( \frac{\partial u'_3}{\partial x_1} \right)^2} + \overline{\left( \frac{\partial u'_2}{\partial x_3} \right)^2} + \overline{\left( \frac{\partial u'_3}{\partial x_2} \right)^2}$$
$$\left. + 2 \left( \overline{\frac{\partial u'_1}{\partial x_2} \frac{\partial u'_2}{\partial x_1}} + \overline{\frac{\partial u'_1}{\partial x_3} \frac{\partial u'_3}{\partial x_1}} + \overline{\frac{\partial u'_2}{\partial x_3} \frac{\partial u'_3}{\partial x_2}} \right) \right\}. \quad (A\,1)$$

The sixth-order polynomial used for the inlet velocity profile is

$$V = -2.63t^6 + 4.84t^5 - 3.42t^4 + 1.17t^3 - 0.20t^2 + 0.02t,$$

where $t$ is the simulation time in seconds and $V$ is the velocity imposed on the hemisphere in ms$^{-1}$.

Turbulence intensity ($T_u$) values, defined as

$$T_u = \frac{\sqrt{\frac{1}{3}(\overline{u'^2} + \overline{v'^2} + \overline{w'^2})}}{V_{mag}} = \sqrt{\frac{1}{3}} \frac{V'_{mag}}{V_{mag}},$$

where $u'$, $v'$ and $w'$ are the fluctuating components of velocity, are reported along with those for $V_{mag}$ and $V'_{mag}$ in table 1. Hahn et al. [39] reported experimentally determined values of $T_u$ in the range 1.5–5% at point locations in the nose. This measure is subjected to large variations in regions with low mean flow values, however.

The simulation was repeated on a coarse mesh with approximately one-fifth as many elements and at a time step 10 times larger. The partitioning of mass flow between the various sectors lacked only the minor high-frequency oscillations visible in figure 8, but was otherwise virtually identical. Likewise, the computed pressure loss between each plane in figure 3 differed between the simulations by less than 4% (ignoring the high-frequency fluctuations) for nearly all planes. The exception was the transglottal loss (planes 8–10), where the coarse simulation recorded a 9% lower pressure drop. Because the latter region contains the main laryngeal jet and thus the most chaotic flow, it could be expected that incomplete resolution would be most evident in this zone.

Scalar arrival times at the carina were the same in both simulations and the temporal concentration distributions matched to within approximately 3%.

Simulations at lower resolution thus appear capable of capturing the overall dynamics of the flow, with adequate accuracy given the multiple physiological uncertainties. However, for measures such as heat and water exchange and particle dispersion that are more sensitive to flow gradients, it is less likely that the coarse resolution would be adequate, without some form of additional modelling.